\documentclass[10pt]{article}
\usepackage{latexsym,graphicx}
\usepackage{latexsym,graphicx}
\newcommand{\be}{\begin{equation}}
\newcommand{\ee}{\end{equation}}
\def\n{\noindent}
\catcode `\@=11 \catcode `\@=12
\begin{document}
\begin{center}
\large{\bf {A Novel Concept for Mass as Complex-Mass towards Wave-particle Duality}} \\
\vspace{10mm}
\normalsize{R. C. Gupta $^1$,  Anirudh Pradhan $^2$ and Sushant Gupta $^3$} \\
\vspace{5mm}
\normalsize{$^{1}$ Institute of Technology (GLAITM), Mathura-281 406, India\\
E-mail: rcg\_iet@hotmail.com, rcgupta@glaitm.org}\\
\vspace{5mm} \normalsize{$^{2}$ Department of Mathematics, Hindu
Post-graduate College, Zamania-232 331, Ghazipur, India \\
E-mail: acpradhan@yahoo.com, pradhan@iucaa.ernet.in}\\
\vspace{5mm} \normalsize{$^{3}$ Department of Physics,
University of Lucknow, Lucknow-226 007, India}\\
\normalsize{$^{3}$ E-mail: sushant1586@gmail.com}\\
\end{center}
\vspace{10mm}
\date{Oct 30, 2009}
\begin{abstract}
In the present paper a new concept is introduced that: `mass is a
complex quantity'. The concept of complex-mass suggests that the
total mass M of a moving body is complex sum of: (i) the real-part
(grain or rest) mass $m_{g}$ establishing its particle behavior and
(ii) the imaginary-part mass $m_{p}$ governing its wave properties.
Mathematically, the complex mass $M = m_{g} + i m_{p}$; the
magnitude $\mid M \mid$ = ($m_{g}^{2}$ + $
m_{p}^{2})^{\frac{1}{2}}$. The theory proposed here explains
successfully several effects such as `Compton effect' and
`refraction of light' which could not be explained otherwise by a
single theory of wave or particle. Also explained are `Doppler
effect for light', `photo-electric effect', `Uncertainty principle',
`Relativity' and `supersymmetry'.
\end{abstract}
\smallskip
\n Key words: Complex mass, Wave particle duality, Photon scattering,
Compton effect, Relativity.  \\
\n PACS: 03.65.-w, 34.50.-s, 03.30.+p \\
\section{Introduction}
The longest controversial issue in the scientific history is perhaps
over the fundamental question of Nature that whether light is wave
or particle? Huygens' wave-theory succeeded over Newton's
corpuscular theory, as the wave theory explained successfully
several important phenomena such as refraction and interference
whereas particle theory could not. However, in the beginning of 20th
century the particle-theory of light (photon) emerged again as it
explained clearly some observations such as photoelectric effect and
Compton effect whereas the wave-theory failed to do so. In order to
resolve the controversy, de-Broglie \cite{ref1} proposed the
hypothesis for wave-particle duality suggesting that light is both,
particle and wave. Moreover he also suggested that a moving electron
too can exhibit wave properties and it was experimentally proved to
be so. However, the wave-particle dilemma especially for light still
persists \cite{ref2,ref6} in some ways. It is not cleat as to why
electromagnetic radiation behaves as particle in one experiment and
as wavy in another experiment. Both the particle and wave aspects
have never been observed simultaneously, as if one aspect comes into
being only if the other aspect is absent and vice-versa. Attempt to
identify one aspect vanishes the other \cite{ref2,ref6}. In an
attempt to resolve the dilemma, the novel concept of `complex mass'
is presented, and its applicability to several situations are
demonstrated in this paper.

\section{Need for the new concept of mass}
Many arguments could be given in favour of the need for the new
concept of mass.
However, the author would state/quote a few, as follows: \\
\\(i) Mass and energy are no more separate quantities, but are
interwoven by Einstein's mass-energy equation. This may imply that
mass losses its independent identity and requires a new
interpretation.\\
\\(ii) It is known, as Ugarov \cite{ref7} mentions in his book on
special relativity that `rest mass of a system exceeds the sum of
rest masses of constituent particles by a certain amount estimated
in the reference frame in which the total momentum is zero'. He
states further that `rest mass is not additive quantity'. Such a
property of mass is uncommon in classical mechanics.
It is tempting to bring in some new definition of mass for constituent
particle.\\
\\(iii) Is mass a scalar-real-quantity, a vector-imaginary-quantity or
a complex-quantity? Rest mass is definitely a real physical
quantity. Photon however has a zero real-physical rest-mass though
it has a total mass ($h\nu/c)/c$ due to its momentum vector
$h\nu/c$, as if it is due to the mass residing therein as
imaginary-part. What about the mass of a moving body having a
rest-mass and momentum-vector? The quaternion was first proposed by
Hamilton \cite{ref8} as a sum of scalar plus vector [8 - 10]; is
mass `a quaternion' for a moving body? Yes, it seems to be so
\cite{ref11}. Alternatively; the representation of a moving mass as
`a complex-quantity' is more meaningful for its familiarity and ease
of writing \& comprehension, and that is what is done in the present
paper as a better way than the earlier paper \cite{ref11}. Anyway,
the mathematics [8 - 10] of `quaternion' and `complex number'are
almost equivalent.\\
\\(iv) Both the aspects of a moving body i.e., the corpuscular
(material) nature and its wave behavior have never been observed
simultaneously. Any attempt to identify one aspect vanishes the
other aspect, as if these two aspects come out of two different
entities (say, real scalar-part and imaginary vector-part). Could
mass be considered as `complex quantity', the real part of it
representing the material-content (rest-mass), and the imaginary
part (due to its momentum) governing its wave aspect? \\
\\(v) If introduction of a new concept of mass as `complex quantity'
could explain various phenomena and could eliminate/minimize
ambiguity; it is worth doing. The new concept may appear to be a bit
speculative at first glance, but as it will be seen later that its
success and capability to explain several diverse phenomena is
striking. All avenues for the `Truth' must be kept open.
\section{Complex-mass concept}
For the `complex mass concept' introduced here it is proposed that
`mass is a complex quantity' and that the total mass M of a moving
body has two components: (i) a real-part (particle at
relatively-rest) `grain-mass' $m_{g}$ and (ii) an imaginary-part
`photonic mass' $m_{p}$ due to its momentum. This may tentatively be
considered as a postulate for the time being. However, it will be
evident later that this novel concept of the `complex-mass' has the
potential to explain several phenomena without any real
contradiction with the present status of formulation in physical
sciences of concern.\\
\\The `total complex-mass' M could be written as the complex sum of
$m_{g}$ and $m_{p}$ as follows:
\begin{equation}
\label{eq1} M = m_{g} + i m_{p}.
\end{equation}
When the particle-momentum is zero (particle at rest) its photonic
(imaginary-part) mass is zero, whereas if its velocity is c (as for
photon) its rest or grain (real-part) mass is zero. Taking x-axis
corresponding to the rest or gain mass and the y-axis corresponding
to the photonic mass, a diagrammatic representation of complex mass
is suggested in Fig. 1.\\
\begin{figure}[htbp]
\centering
\includegraphics[width=13cm,height=12cm,angle=0]{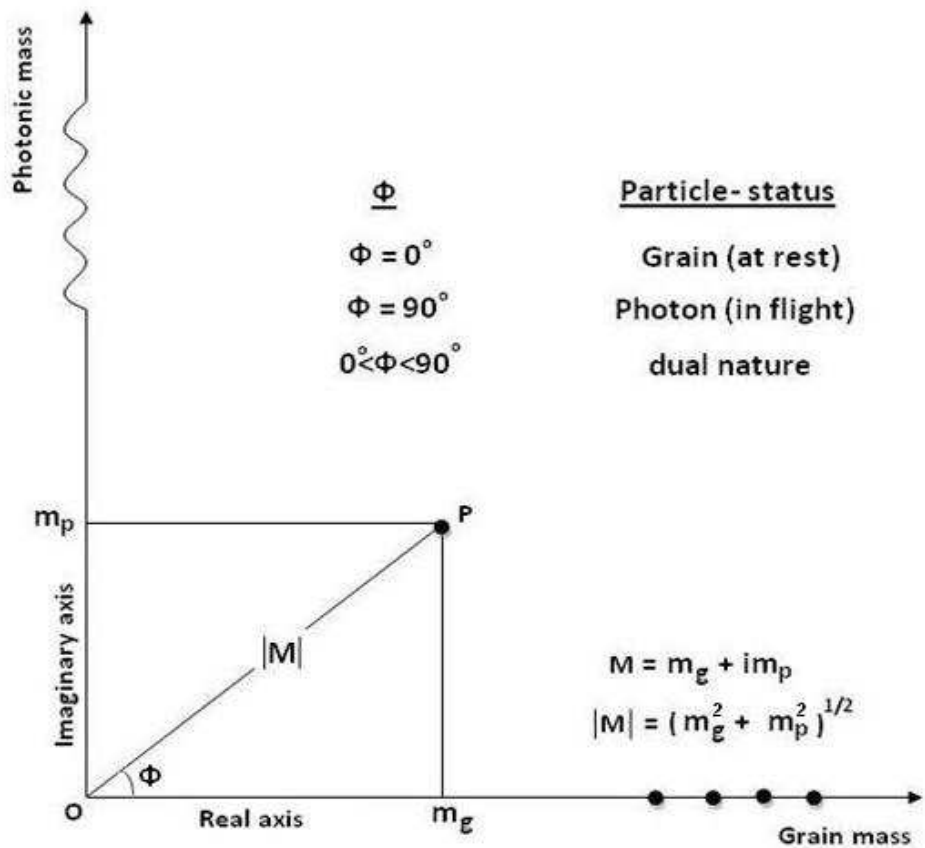}
\caption{Complex-mass representation}
\end{figure}
\\Total `magnitude' $\mid M \mid$ of the complex-mass M is
`Pythagorean-sum' of the constituents as follows as per property of
the complex-number,
\begin{equation}
\label{eq2} \mid M \mid = (m_{g}^{2} + m_{p}^{2})^{\frac{1}{2}}.
\end{equation}
and the `grain-photon phase angle' $\phi$ (Fig.1) is defined as,
\begin{equation}
\label{eq3} \cos\phi = \frac{m_{g}}{\mid M \mid} and \sin\phi =
\frac{m_{p}}{\mid M \mid}.
\end{equation}
Though the `grain-photon angle $ \phi $' is different from the
`Schrodinger's wave-function $ \psi $', but seems to have some
correlation; greater value of $ \phi $ implies that the particle is
more `wavy' than `grainy' i.e., it has more wave-aspects.\\
\\Particle's relative speed is v and photon's speed is c. The `total'
momentum of the particle is $ \mid M \mid v $. It is considered that
the rest (real-part) mass does not take any momentum and the total
momentum is taken up by the photonic (imaginary-part) mass ($
m_{p}$). Hence, $ \mid M \mid v = m_{p}c $ or
\begin{equation}
\label{eq4} m_{p} = \mid M \mid \frac{v}{c}.
\end{equation}
From equations (2) and (4) the following equation is obtained,
\begin{equation}
\label{eq5} \mid M \mid  = \frac{m_{g}}{\left(1 -
\frac{v^{2}}{c^{2}}\right)^{\frac{1}{2}}},
\end{equation}
which may also be rewritten as
\begin{equation}
\label{eq6} m_{g} = \mid M \mid \left(1 -
\frac{v^{2}}{c^{2}}\right)^{\frac{1}{2}}.
\end{equation}
The cross-relationship between the magnitude-of-mass $ \mid M \mid $
and photonic-mass $ m_{p} $ (as in Eqs. 2 \& 4) may appear to be
interestingly intriguing showing dependence of one on the other, but
is okay since it includes the relativistic effect as reflected in
Eqs. (5) and (6). \\
\\ If $\phi = 0^{0}$, the particle is a real-particle (grain at rest) as in
Fig.1; if $\phi = 90^{0}$, the particle is in-particle-sense an
imaginary-particle (photon in flight). For a particle moving with a
velocity $(v <c)$, $ 0^{0} < \phi < 90^{0} $ and it has dual
(complex) nature. Mathematically (from Eqs. 1, 4, 5 \& 6),
\begin{equation}
\label{eq7} M = m_{g} + im_{p} = m_{g} + i \mid M \mid \frac{v}{c} =
\mid M \mid \left[\left(1 - \frac{v^{2}}{c^{2}}\right)^{\frac{1}{2}}
+ i \frac{v}{c}\right] .
\end{equation}
\subsection{de-Broglie Wavelength}
Photonic-momentum $ p ( = m_{p} c = \mid M \mid v) $ and
photonic-energy $ E ( = h \nu)$ are as usual related by $ E = p c =
\mid M \mid v c $. Rest-mass does not take any momentum, total
momentum is taken up by photonic mass $ m_{p} $ leading to the
de-Broglie wavelength $ \lambda $ (Eq. 8) as under; as $ p = \mid M
\mid v = m_{p} c = \frac {E}{c} = \frac{h\nu}{c}$ and that $ c =
\nu\lambda $ , where h is the Planck's constant and $\nu$ is the
frequency, $\lambda$ is the wavelength and c is the speed of light,
\begin{equation}
\label{eq8} \lambda = \frac{h}{(\mid M \mid v)}.
\end{equation}
\subsection{Complex-Mass of a moving Particle and The Two Ways of
its Transformation:}
\subsubsection{Self-inspired self-conversion of rest-mass to photonic-mass
without any external agency} Consider a particle at rest (point A in
Fig. 2a) having only the material-like grain-mass and no wave-like
photonic-mass. So, total mass $ \mid M \mid = m_{g} $. Now think
that this stationary particle suddenly (self-inspired) starts moving
at velocity $v$ (point $A^{\prime}$ in Fig. 2b) thus gaining
photonic mass $ m_{p} = \mid M \mid \frac {v}{c}$ at the expense of
its own grain-mass which is reduced to a new lower value $
m_{g}^{\prime}$; total mass $\mid M \mid$ remains same as no
external agency is involved in it. Complex mass (for Fig. 2b) is
thus expressed as
\begin{figure}[htbp]
\centering
\includegraphics[width=13cm,height=8cm,angle=0]{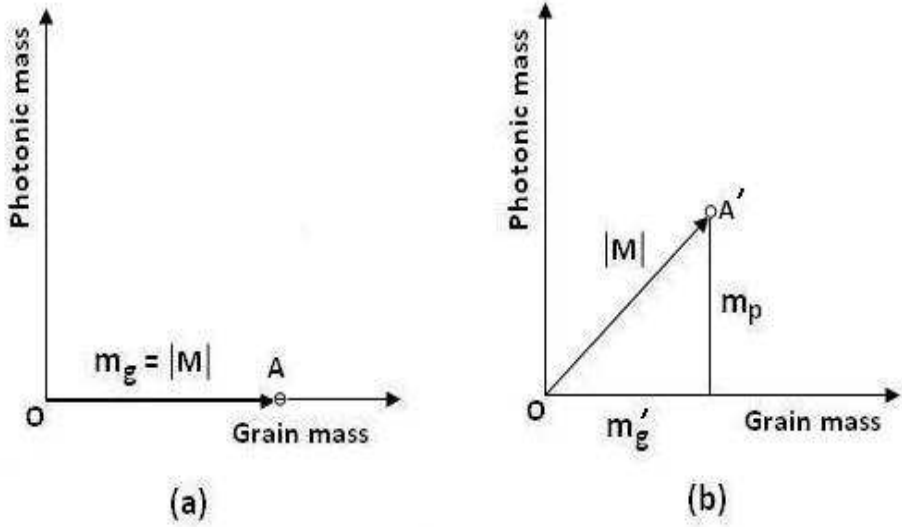}
\caption{Self-inspired self-conversion of rest mass to photonic
mass}
\end{figure}
\begin{equation}
\label{eq9} M = {m_{g}}' + im_{p}.
\end{equation}
Magnitude wise, $$ \mid M \mid ^{2} = {{m_{g}}'} ^ {2} + m_{p} ^ {2}
= {{m_{g}}'}^{2} + ( \mid M \mid \frac {v} {c} )^{2}$$ or
\begin{equation}
\label{eq10} {m_{g}}' = \mid M \mid \left(1 -
\frac{v^{2}}{c^{2}}\right)^{\frac{1}{2}} = m_{g} \left(1 -
\frac{v^{2}}{c^{2}}\right)^{\frac{1}{2}},
\end{equation}
\mbox{since} from Fig. (2a) $\mid M \mid = m_{g}$. Note that $m_{g}
$ is the (rest) grain-mass of a stationary-particle whereas $
m_{g}^{\prime} $ is the reduced grain-mass of the self-inspired
moving particle, also note that in the case of self-inspired
self-conversion of (rest) grain-mass, the grain-mass decreases (from
$m_{g} $ to $ m_{g}^{\prime} $) as the particle moves faster.
Equations (10) and (6) are similar. Self-conversion of grain-mass to
photonic-mass and vice-versa are manifested fully in `annihilation'
and `pair-production' processes for elementary
particles and antiparticles.\\
\\The total mass M of the particle when $ v = 0 $ is $ \mid M \mid $,
from Eq. (7), stationary grain-mass is $ \mid M \mid $ and
photonic-mass is zero. If, however, the stationary mass converts
itself (self-inspired) into photonic-energy (wave) it may be
considered as moving with a photonic-speed ($v = c$) with grain or
rest mass as zero and photonic mass as $ \mid M \mid $ leading to
Einstein's mass energy equation (Eq. 11) as photonic energy $ E =
\mid M \mid v c $  as mentioned in section 3.1,
\begin{equation}
\label{eq11} E = \mid M \mid c^{2}.
\end{equation}

\subsubsection{Induced-motion due to addition of energy to the
particle by external-agency} Consider a particle at rest (point A in
Fig. 3a) having only a stationary grain-mass $ m_{g} $ with total
mass $ \mid M \mid  = m_{g} $. Now consider that this mass is pushed
or worked-upon or given energy to move forward with a velocity $v$.
Thus a photonic-mass is added (Pythagorean way) onto the grain-mass
(Fig. 3b) due to the work or energy supplied to it. The total mass
thus increases to a higher value $ M^{\prime} $, given by, as
complex sum as,
\begin{equation}
\label{eq12} M' = m_{g} + i m_{p}.
\end{equation}
\begin{figure}[htbp]
\centering
\includegraphics[width=13cm,height=8cm,angle=0]{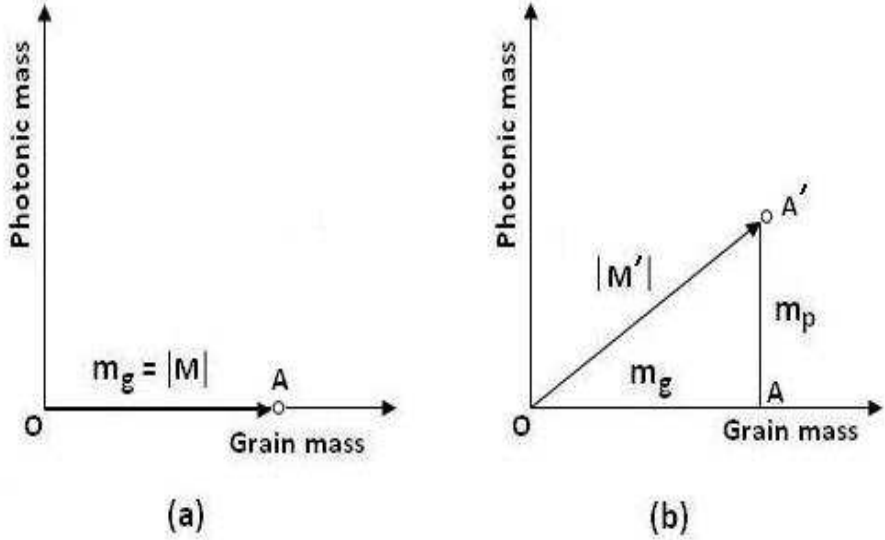}
\caption{Pythagorean addition of energy $m_{g}$ (by external agency)
to the particle as photonic mass }
\end{figure}
Magnitude wise, $$ \mid M^{\prime} \mid ^{2} = m_{g} ^ {2} + m_{p} ^
{2} = m_{g}^{2} + \left( \mid M^{\prime} \mid \frac {v} {c}
\right)^{2}$$ or
\begin{equation}
\label{eq13} \mid M' \mid = \frac{m_{g}}{\left(1 -
\frac{v^{2}}{c^{2}}\right)^{\frac{1}{2}}}.
\end{equation}
Thus, there is an overall increase in the total mass from $ \mid M
\mid $ to $ \mid M^{\prime} \mid $ due to addition of energy. Note
that the mass of the stationary-particle is $ \mid M \mid  = m_{g}$,
whereas when-after addition of work or energy $(W = E = m_{p}c^{2})$
the particle gains photonic mass $m_{p}$ it starts moving with
greater total mass $ \mid M^{\prime} \mid $. Equations (13) and (5)
are similar. Also note that the Eqs. (13 or 5) is similar and
equivalent to the famous Relativity-formula $$ m =
\frac{m_{0}}{(1-\frac{v^{2}}{c^{2}})^\frac{1}{2}},$$ where $m_{0}$
is the rest-mass and m is the total moving mass.

\section{Collision of particles and conservation laws}

Consider two particles 1 and 2, initial masses of which
before-collision are $M_{1} = m_{g1} + i m_{p1}$ and $ M_{2} =
m_{g2} + i m_{p2}$. During collision, the grain and the photonic
masses could be `re-distributed' between them. Consider that
after-collision the new masses for the two particles are
$M_{1}^{\prime} = m_{g1}^{\prime} + i m_{p1}^{\prime}$ and
$M_{2}^{\prime} = m_{g2}^{\prime} +i m_{p2}^{\prime}$.\\
\\The total magnitude of the mass $\mid M \mid $ for the colliding
particles (which takes into account `relativistically' both, the
grain-mass and the photonic-mass) would therefore be `conserved'
giving the following equation,
\begin{equation}
\label{eq14} \mid M_{1} \mid + \mid M_{2} \mid = \mid M_{1}' \mid +
\mid M_{2}' \mid.
\end{equation}
In the process of the collision of the particles, the total vector
(imaginary-part) photonic mass is also considered to be `conserved'
and thus,
\begin{equation}
\label{eq15} \mid m_{p1} \mid + \mid m_{p2} \mid = \mid m_{p1}' \mid
+ \mid m_{p2}' \mid.
\end{equation}
In fact, as it will be more evident later in this paper that the
conservation of total mass $\mid M \mid $ (Eq. 14) is the
mass-energy conservation and that the conservation of photonic mass
$m_{p}$ (Eq. 15) is the conservation of momentum.\\
\\It may be noted that the `complex mass' introduced in this paper
takes into account the relativistic aspects as reflected in Eqs. (5)
- (7). Thus, in general, the real-part i.e., the rest (grain) mass
is not conserved in collision unless the colliding particles retain
their identities. It is also noted that the `complex-mass' is
somewhat different from the usual mathematical complex-number; as
complex-mass is in fact relativistic.

\section{Photon scattering}
To study the photon scattering, take a general case of collision
(Fig. 4) where a photon strikes (at an angle $ \alpha $) on a moving
object (say, electron) and scatters away. Consider that
before-collision the electron moving with a velocity $ v_{1} $ has a
total mass $ M_{1} = m_{g} + i \mid M_{1} \mid \frac {v_{1}}{c}$
where $m_{g} $ is rest-mass of the electron, and that the incident
photon (with a momentum $\frac{h\nu}{c}$) has a total mass $M_{2} =
0 + i \frac{h\nu}{c^{2}}$. After collision; the photon (at an angle
$\theta$, with momentum $h\nu^{\prime}$) is emerged with a total
mass $ M_{2}^{\prime} = 0 + i \frac{h\nu^{\prime}}{c^{2}}$, and that
the electron comes out (at an angle $\delta$) moving with a velocity
$v_{1}^{\prime}$ which has a total mass $ M_{1}^{\prime} = m_{g} + i
\mid M_{1}^{\prime}\mid \frac{v_1^{\prime}}{c}$.\\
\begin{figure}[htbp]
\centering
\includegraphics[width=13cm,height=8cm,angle=0]{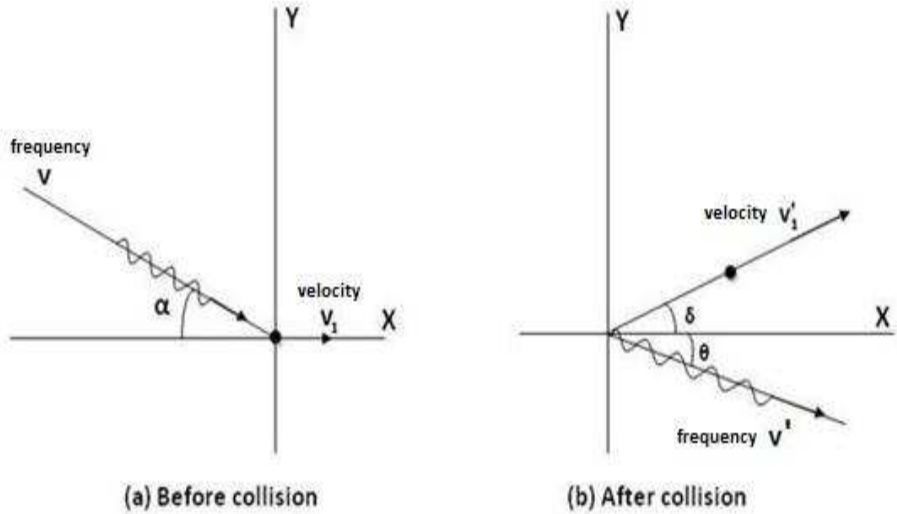}
\caption{Photon Scattering}
\end{figure}
\\Equation (15) for photonic-mass conservation reduces to the
following equations (Eqs. 16 \& 17) of  conservation corresponding
to the momentum conservation in x and y directions,
\begin{equation}
\label{eq16} \mid M_{1} \mid v_{1} + \frac{h\nu \cos{\alpha}}{c} =
\mid M_{1}' \mid v_{1}' \cos{\delta} + \frac{h\nu'\cos{\theta}}{c},
\end{equation}
\begin{equation}
\label{eq17} \frac{h\nu}{c \sin{\alpha}} = - \mid M_{1}' \mid v_{1}'
\sin{\delta} + \frac{h\nu'}{c \sin{\theta}}.
\end{equation}
Equation (14) for total-mass conservation reduces to mass-energy
conservation equation (Eq. 18) as,
\begin{equation}
\label{eq18} \mid M_{1} \mid c^{2} + h \nu = \mid M_{1}' \mid c^{2}
+ h \nu'.
\end{equation}
where from Eq. (5) magnitude of total  complex masses of the
electron before and after the collision (by photon) are as follows,
\begin{equation}
\label{eq19} \mid M_{1} \mid = \frac{m_{g}}{\left(1 -
\frac{v_{1}^{2}}{c_{1}^{2}}\right)^{\frac{1}{2}}} ~ ~ \mbox{and} ~ ~
\mid M_{1}' \mid = \frac{m_{g}}{\left(1 -
\frac{v_{1}'^{2}}{c_{1}^{2}}\right)^{\frac{1}{2}}}.
\end{equation}
From Eqs. (16), (17), (18) and (19), the following expression is
obtained,
\begin{equation}
\label{eq20} (\nu - \nu') = \frac{(h \nu \nu'\cos{\alpha})}{(m_{g}
c^{2})}[1 - \cos{(\theta - \alpha)}] + \frac{v_{1}}{c}(\nu
\cos{\alpha} - \nu'\cos{\theta}).
\end{equation}
Laugier \cite{ref12} gave a more accurate derivation of above Eq.
(20) as follows, however, as he himself mentioned that the slight
correction has no consequence for the rest of the paper.
\begin{equation}
\label{eq21} (\nu - \nu') = \frac{h \nu \nu'}{m_{g} c^{2}}\left(1 -
\frac{\nu_{1}^{2}}{c^{2}}\right)[1 - \cos{(\theta - \alpha)}] +
\frac{v_{1}}{c}(\nu \cos{\alpha} - \nu'\cos{\theta}).
\end{equation}
For cases where the incident photon strikes not the electron but
some `heavier' object such as an observer (or an observing
instrument), mirror or glass; $m_{g}$ will be replaced with
rest-mass of the object in Eq. (20) where the first term on R.H.S.
would become negligible and the equation would thus reduce to,
\begin{equation}
\label{eq22} (\nu - \nu') = (\nu \cos{\alpha} - \nu'
\cos{\theta})\frac{v_{1}}{c}.
\end{equation}
\subsection{Compton Effect}
For $v_{1} = 0$ and $\alpha=0$, the Eq. (20) reduces to the equation
for Compton effect \cite{ref13} as follows,
\begin{equation}
\label{eq23} \left(\frac{1}{\nu'} - \frac{1}{\nu}\right) =
\frac{h}{(m_{g} c^{2})}(1 - \cos{\theta}).
\end{equation}
\subsection{Doppler Effect (For Light)}
If the incident light (at $\alpha$ = 0) reflects back (at $\theta =
180^{0}$) from the moving object (mirror), the Eq. (22) reduces to
the following, after some algebraic manipulations,
\begin{equation}
\label{eq23} \frac{\nu'}{\nu} = \frac{\left(1 -
\frac{v_{1}}{c}\right)}{\left(1 + \frac{v_{1}}{c}\right)}.
\end{equation}
which gives the frequency of the reflected light ($\nu^{\prime}$)
from an object - which is receding from the source (which emits
light of frequency $\nu$). It should be noted that this frequency
($\nu^{\prime}$) is Doppler-shifted twice, firstly when the moving
object (mirror) receives the light and secondly when it reflects.
The result obtained here is consistent with that predicted by
Doppler-effect \cite{ref14} for light.
\subsection{Reflection and Refraction}
\subsubsection{Reflection}
Consider that light of frequency v strikes a stationary mirror
$(v_{1} = 0)$ and that the frequency of the reflected light is
$v^{\prime}$. From Eq. (22), $v^{\prime} = v$. Moreover, since free
surface has zero shear-stress, momentum along free-surface should be
same before and after impact. Mathematically,
$$\frac{h\nu}{c}\sin\alpha = \frac{h\nu^{\prime}}{c}\sin\beta$$ gives
the reflection-law \cite{ref15} as follows (incident angle $\alpha$
equals reflected angle $\beta$), since $ \nu = \nu^{\prime}$,
\begin{equation}
\label{eq25} \alpha = \beta.
\end{equation}
\subsubsection{Refraction}
As in the case of reflection; for refraction of light into a
stationary $(v_{1} = 0)$ transparent medium, Eq. (22) gives
$\nu^{\prime} = \nu$. In case of reflection both the incident and
reflected light travel in the same medium, but for refraction it is
different. For refraction case, the incident light of frequency
$\nu$ travels in a medium(say, air) wherein speed of light is c and
the refracted light of frequency $\nu^{\prime}$ travels in another
medium(say, glass) wherein speed of light is $c^{\prime}$. Thus the
momentum of the incident light is $\frac{h\nu}{c}$ whereas the
momentum of refracted light is $\frac{h\nu^{\prime}}{c^{\prime}}$.
Here again, since the interface between the two media can-not take
shear stress, momentum conservation along the interface must hold
good i.e., $$\frac{h\nu}{c}\sin\alpha =
\frac{h\nu^{\prime}}{c^{\prime}}\sin\gamma,$$ which gives the
Snell's law \cite{ref15} for refraction as follows, since $ \nu =
\nu^{\prime}$ ,
\begin{equation}
\label{eq26} \frac{\sin{\alpha}}{\sin{\gamma}} = \frac{c}{c'} = \mu,
\end{equation}
where $\mu$ is the refractive index of the second medium-material
(say, glass $\mu > 1$) with respect to the first medium (say, air),
which indicates that speed of light ($c^{\prime}$) in denser medium
(glass) would be less than the speed of light (c) in
rarer-medium(air), as expected and is in accordance with the results
of wave theory.\\
\\It is to recall that the main reason for failure of Newtonian
corpuscular-theory was that it predicts wrong results about speed of
light in denser medium whereas the wave-theory predicts correctly.
The complex-mass concept, which considers light as particle of
photonic-mass (which differs in the two media, $\frac{h\nu}{c^{2}}$
in one medium and $\frac{h\nu^{\prime}}{{c^{\prime}}^{2}}$ in the
other medium), also predicts (Eq. 26) rightly.

\subsubsection{Partial refraction and partial reflection}

Consider that the incident light (n photons) strikes the stationary
($v_{1}=0$) glass and that it is partially transmitted ($n_{2}$
photons) into and partially reflected ($n_{1}$ photons) back, such
that $n = n_{1} + n_{2}$. As shown earlier, the frequencies: of
incident ($\nu$), transmitted ($\nu_{t}^{\prime}$) and reflected
($\nu_{r}^{\prime}$) light are same. Here also, since the
free-surface can-not take shear-stress, momentum conservation along
the interface yields,
$$ n\left(\frac{h\nu}{c}\right)\sin{\alpha} = n_{1}\left(\frac{h\nu_{r}'}
{c}\right)\sin{\beta} + n_{2}\left(\frac{h\nu_{t}'}
{c'}\right)\sin{\gamma},$$ where $\alpha$, $\beta$ and $\gamma$ are
angle(s) of incident, refraction and reflection respectively. \\
\\Taking $n = n_{1} + n_{2}$ and $\nu$ = $\nu_{t}^{\prime}$ =
$\nu_{r}^{\prime}$ the above equation reduces to
\begin{equation}
\label{eq27} n_{1}[\sin{\alpha} - \sin{\beta}] +
n_{2}\left[\sin{\alpha} -
\left(\frac{c}{c'}\right)\sin{\gamma}\right] = 0.
\end{equation}
For partial refraction and partial reflection, $n_{2}$ and $n_{1}$
are the positive integers thus Eq. (27) leads to both, the
reflection-law (Eq. 25) and the refraction-law (Eq. 26). Possibly,
specific combinations of $n_{1}$ and $n_{2}$ could lead to the
understanding of bunching / anti-bunching of light \cite{ref16}.
\section{Photo-electric effect}
When a photon of energy $(h\nu)$ strikes an electron (in an atom),
part of its energy (W, the work-function) is used up for removing
the electron from the atom, thus the remaining energy $(h\nu - W)$
is used to give a velocity v to the electron.\\
If $ m_{e} $ is the total mass of an electron in the atom,
conservation of total mass (Eq. 14) gives,
$$ m_{e} + \frac{(h\nu - W)}{c^{2}} = \frac{m_{e}}{\left(1 -
\frac{v^{2}}{c^{2}}\right)^{\frac{1}{2}}} + 0,$$ which gives the
following Einstein's equation (28) for photo-electric effect
\cite{ref17} as follows (neglecting the higher order terms),
\begin{equation}
\label{eq28} h\nu = \frac{1}{2}m_{e}v^{2} + W.
\end{equation}
Photon during its flight actually behaves as wave, but when it
strikes an objects and it stops (either momentarily as in
Compton-effect or completely as in photo-electric effect) the photon
exhibits particle-like aspect.

\section{Uncertainty Principle}
Consider (in Fig.4) that a photon of frequency $\nu$ strikes (at
$\alpha = 0$) an electron and reflects back (at $\theta = \pi $)
with a frequency $\nu^{\prime}$ (from Eq. (23) $\nu$ may be
different from $\nu'$ but as h is very small, $\nu^{\prime} \approx \nu$).\\
\\It can be shown from Eq. (16) that the magnitude of change
(uncertainty) in the momentum (for $\alpha = 0 = \delta, \theta =
\pi $)  $\triangle{mv} = 2 \frac {h\nu}{c}$.\\
\\Position of the electron from the photon-source is x. Distance
traveled by the emitted photon to return back is 2x. During this
distance it may be considered that the light had N wavelengths or $
2x \approx N \lambda $. The uncertainty in the measurement of
position (or in N), would be of the order of an error of one
wavelength, thus $\triangle{x} \approx \frac {\lambda}{2}$.
Therefore,
\begin{equation}
\label{eq29} \triangle{mv} \triangle{x} \approx h,
\end{equation}
which is of the same order of magnitude as that by the famous
Heisenberg's uncertainty principle \cite{ref18}.\\
\\Here not only that it is shown that product of uncertainty in
momentum and position is of the order of h but also shown
individually that error in position is half the wavelength and that
error in momentum would be twice the photon-momentum. For
measurement with high frequency (small $\lambda$) photons,
$\triangle x$ would be less but $\triangle mv $ would be more,
product of these $\approx h$.  A more rigorous consideration
(scattering in all directions) would result this product as $\frac
{h}{(2\pi)}$.

\section{Relativity}
From Eqs. (2) and (4) the relativistic energy equation could be
written as follows,
\begin{equation}
\label{eq30} \mid M \mid^{2}c^{4} = m_{g}^{2} c^{4} + \mid M
\mid^{2} v^{2} c^{2},
\end{equation}
\begin{equation}
\label{eq31} E^{2} = E_{g}^{2} + p^{2}c^{2},
\end{equation}
where total-energy $ E = \mid M \mid c^{2} $, relative-rest-energy
$E_{g} = m_{g}c^{2}$ and photonic-energy $ E_{p} = \mid M \mid v c
$; total momentum = photonic momentum $ = m_{p}c = \mid M \mid v =
p$.\\
\\Considering the momentum vector $$ \mid M \mid v = \mid M \mid (i v_{x}+ j
v_{y}+ kv_{z}) = (i p_{x} + j p{y}+ k p{z}) = p,$$ it could be shown
from Eq. (31) that
\begin{equation}
\label{eq32} \left(\frac{E_{g}}{c}\right)^{2} =
\left(\frac{E}{c}\right)^{2} - (p_{x}^{2} + p_{y}^{2} + p_{z}^{2}) =
\left(\frac{E'}{c}\right)^{2} - (p_{x}'^{2} + p_{y}'^{2} +
p_{z}'^{2})
\end{equation}
which is nothing but the well-known `4-vector invariant' in unprimed
and primed ($\prime$) coordinate system [19 - 21]. The $4-$vector
invariance of momentum is well known in special-relativity.\\
\\It may be noted that the relativistic energy equations (30 - 31)
are derived from complex-mass concept equations (1 to 7). The
concept of `complex mass' finds strength from the fact that reverse
is also possible i.e., the complex-mass concept equations (1 to 7)
could be derived from the relativistic energy equations (30 - 31).\\
\\In fact, the 4-vector invariance (Eq. 32) is equivalent to the
invariance of real-part rest (grain) mass, in unprimed and primed
coordinate system. This is consistent with the concept of complex
mass; because a change of coordinate system would affect velocity to
change the vector (imaginary-part) photonic-mass thus changing the
total-mass, whereas the scalar (real-part) grain-mass (material)
remains unchanged.\\
\\The special-relativity in the concept of complex-mass is also
reflected in Eqs. (5 - 6) and Eqs. (13 \& 10) in fact, the
special-relativity is inherent in the concept of complex mass.

\section{Supersymmetry and Links between Complex-mass and Supersymmetry}

Supersymmetry \cite{ref22} means that there is transformation which
relates the particles of integral spin such as photon (boson) to the
particles of half-integral spin such as electron (fermion). Bosons
are the `mediators' of the fundamental forces while fermions make up
the `matter'. The supersymmetry solves the `hierarchy problem' for
grand-unification. Also, for unification of forces, with
supersymmetry the promising string-theory becomes the better and
famous superstring theory \cite{ref23}. But as yet, no
supersymmetric -partner particles have been found. An ambitious
attempt is made, as follows, to answer: `why superpartners exist
only in {\it principle} but not in {\it reality}'. \\\\
Referring to the `complex mass concept', it may be noted that the
grain-mass $m_{g}$ signifies the material content due to
atoms/molecules (group of fermions) whereas the photonic-mass
$m_{p}$ is due to associated photon (boson). So, in a way, the
complex mass $M = m_{g} + i m_{p}$ is `marriage' of fermion plus
boson. Thus the supersymmetry seems to be inherently embedded in the
complex mass concept. It is as if, fermion \& boson are `coupled to
each other' and that fermion-part behaves as particle \& boson-part
shows the wave nature. Wave-particle duality seems, thus in a way,
due to supersymmetry embedded in the complex mass. It seems that
supersymmetric partners (\& thus the supersymmetry) can exists only
in `married' state as complex sum as $M = m_{g} + i m_{p}$, of
fermionic-part (grain-mass) and bosonic-part (photonic-mass).
\section{Discussions}
It is well known that, as per Relativity-theory, total moving mass
is more than its rest mass, which is also evident from Eq.
(\ref{eq13}) of section-3.2.2. But this is only a part of the story;
the rest (grain) mass itself can decrease if the motion is
self-inspired on its own as given in Eq. (\ref{eq10}) of
section-3.2.1. Note that the equations 5 \& 6 (similar to equations
(\ref{eq13}) \& (\ref{eq10}) respectively) are basically the two
facets of the same coin but have different meanings as explained in
sections-3.2.2 \&
3.2.1.\\
\\To avoid ambiguity of wave-particle, some new terminology is
suggested as follows. A particle at rest (in laboratory) is let
called as `grainon', a particle moving with a speed $v (v < c)$ is
let called as `complexion' and the particle moving with speed of
light is called as `photon'. All; grainon, complexion and photon,
let be `said' as particles! However, the grainon (at relative-rest v
= 0) has corpuscular (material) nature, the photon (in flight v = c)
has the wave nature and the complexion (in motion $v < c$) has both
corpuscular and wave natures. \\
\\However, as per Eq. (\ref{eq11}) inter-conversion of photon and
grainon is possible. Moreover, when a photon is in motion ($v = c$)
it behaves as wave (during its flight), but when it strikes an
object it stops ($v = 0$, momentarily in `Compton-effect' and
completely in `photo-electric effect' experiments) and thus the
photon thereafter behaves as grainon. In experiments such as of
`interference' the two photon-streams interfere in the flight itself
before coming to rest on screen, thereby showing the wave-nature,
the essential-characteristic necessary for such experiments. That's
why photon shows sometimes (in flight) wavy-aspect and sometimes
(when it strikes another particle) corpuscular-nature.\\
\\Furthermore, an atom at rest may be apparently considered as
grainon but it is composed of complexion (moving electron) and
grainon (stationary nucleus), the nucleus in turn is composed of
several particles of varied nature. In a way, one can say that an
atom is a `compound' particle. \\
\\The concept of complex-mass is new, interesting, promising and is
of fundamental importance. Its compatibility with special-relativity
suggests that it is okay but at this stage there is no point in
weighing this (present theory) with full grown 4-vector
special-relativity theory. However, a comparison could be made
between the two for clarity as follows in Table-1.
\newpage
{\bf{Table-1 Comparison between `Complex-mass approach' and
`Special-Relativity theory' for mass.}}\\
\begingroup
\halign{#\hfil& \quad\hfil#& \quad\hfil#&
                \quad\hfil#& \quad\hfil#& \quad\hfil#\cr
                \noalign {\smallskip\hrule\smallskip}
\bf Items&
\hfill\bf Complex-Mass  \hfill&   
\bf Special-Relativity& \hfill\bf Remarks\hfill \cr \noalign
{\smallskip\smallskip}
\bf &   
\hfill\bf Approach \hfill& \hfill\bf Theory\hfill& \hfill\bf
\hfill\cr \noalign {\smallskip\hrule\smallskip}

\cr \bf Masses \\\\\\
&
               \\&
     \cr
Grain(rest)mass  \\ &
            \\ &   \\&    \\&\cr
(real-part)\\ &
            $m_{g}$\\\\ & $m_{0}$  \\\\ & $m_{g}$ = $m_{0}$\\& \cr
\cr Photonic mass    \\ & \\
&
            \\ & \\&     \cr
(imaginary-part)\\ &
            $ m_{p}$ = $\mid M \mid$ $\frac{v}{c}$\\\\ & -  \\\\
            & $ m_{p} =({m}^{2}-{m_{0}}^{2})^{1/2}$\\& \cr
\cr Total magnitude of mass\\ &
            $\mid M \mid$\\\\ & m  \\\\ & $\mid M \mid$ = m\\& \cr

\cr \bf Momentum      \cr

Total momentum    \\ &
            $ \mid M \mid v$\\\\ & $mv$\\\\ & \cr
\cr Photonic momentum    \\ &
            $m_{p}$c = $\mid M \mid v$\\\\ & -\\\\ & \cr

\cr \bf Energy\\  \cr Total energy    \\ &
            $\mid M \mid$ ${c}^{2}$\\\\ & E = m ${c}^{2}$\\\\ & \cr
\cr Rest-mass energy    \\ &
            $m_{g}$ ${c}^{2}$ \\\\ & $E_{0} =$ $m_{0}$ ${c}^{2}$\\\\ & \cr
\cr Photonic energy    \\ &
            $ \mid M \mid vc$\\\\ & -\\\\ & \cr
\cr Kinetic energy\\ &
            - \\\\ & m ${c}^{2}$ - $m_{0}$ ${c}^{2}$ \\\\ &
            $\approx \frac{1}{2}mv^{2}$\\& \cr
\cr \bf Conservation laws of\\
 \cr Photonic mass\\ &
            $m_{p}$ \\\\ & momentum\\\\ & $m_{p} c = mv$ \\& \cr
\cr Total mass \& energy\\ &
            $\mid M \mid $ \\\\ & mass-energy\\\\ &
            $\mid M \mid$ ${c}^{2}$ = m ${c}^{2}$\\&\cr
 \cr \bf Invariance of
 \cr real-part of mass (energy)\\ &
            grain(rest) mass \\\\ & 4-vector momentum\\\\
            &both are equivalent\\&\cr

\cr \bf Representation    \\
 \cr Diagram    \\ &  Complex-mass \\ &
            Minkowskian space\\\\ &    \\ &       \cr
     \\ &  representation\\ &
\cr Dimensions\\ & Real-part scalar(1) + \\ &
            space(3) + time(1)\\\\ &   both, 4-dimensions \\        \cr
\\ & imaginary-part vector(3)\\ &

\cr \bf Expression    \\
 \cr Equation\\ &  $M = m_{g} + i  m_{p}$\\ &
            $(\frac{E_{0}}{c})^{2} = (\frac{E}{c})^{2} - $\\\\
            &   Inter-derivable \\        \cr
  \\ &  \\ &
            $(p_{x}^{2} + p_{y}^{2} + p_{z}^{2})$\\\\ &   \\
            \cr
\noalign {\smallskip\hrule\smallskip}\cr}
\endgroup

The authors, however, like to stress that the complex-mass
components are not merely a re-arrangement of relativistic masses
but the important point is the introduction of the concept that mass
of a moving body is a `complex quantity' i.e., it has two
components: (i) a scalar (real-part) component as grain-mass
(exhibiting the corpuscular material nature) and (ii) a vector
(imaginary-part) component as photonic-mass (governing the wave
aspects). It may also, however, be noted that the `complex mass' is
not simply a conventional type of mathematical complex-number but is
a new type of complex-quantity encompassing the relativistic aspects
too, as reflected in Eq. (\ref{eq7}).\\
\\Although several applications of complex-mass have already been
considered earlier in some depth, the author would cite the
following example where the importance of photonic (vector
imaginary-part) mass may possibly be further emphasized and
usefulness of the complex-mass concept may be appreciated.\\
\\In quantum mechanical explanation of interference pattern of
electrons in the famous double-slit experiment, we are forced to
accept that `an' electron passes through `both' the slits
simultaneously! Whereas in view of the `complexion' concept of
complex-mass; it may be considered that the electron (rest scalar
mass) just-pass through one of slits only and the associated wave
(vector imaginary-part photonic-mass) split-passes through both the
slits, or in other words the `complexion' divided into two parts, or
the `complexion' momentarily divides into two different complexions.
\\
\\ As discussed in section-9, it seems that the supersymmetry is
embedded in coupled-state
into the complex-mass.\\
\\In the complex-mass concept, in fact, both the corpuscular and the
wave aspects are merged together. The concept may provide a
bridge-link between quantum- mechanics and special-relativity as
well as between micro world and macro world. Special theory of
relativity has close links to Electro-magnetism \cite{ref24}.
General theory of relativity \cite{ref25} is the theory of Gravity,
which tells that gravity is there because the 4- dimensional
space-time is curved and that this curvature is because of presence
of `mass' there. Even the alternative theory \cite{ref26} of gravity
suggests that gravity is due the material mass. `Mass' is in the
central theme everywhere from atom to galaxy. What if the mass is a
`complex' quantity; the complex-mass is in accordance with
special-relativity but how is it with general-relativity? Although
at present stage it may appear that the concept of `complex' mass
raises more questions than it provides answers, but the concept is
of fundamental importance and its full potentials will be realized
in time to come.
\section{Conclusions}
The concept of `complex-mass' is new, simple and useful. It has the
mathematical ingredients and the special-relativity is naturally
embedded in it. It is able to explain phenomena which could not
otherwise be explained by a single theory of wave or of particle.
The wave-particle duality is mathematically incorporated in the
complex mass. The novel concept that `mass is a complex quantity' is
of fundamental importance and has the potential to explain several
diverse phenomena. It easily explains (derives) de-Broglie
hypothesis, thus it has ability to minimize the quantum-mystery.
\section*{Acknowledgements}
The author (R. C. Gupta) thanks IET/UPTU, Lucknow and GLAITM,
Mathura for providing facility for the research.

\end{document}